\documentclass[pre,twocolumn,showpacs,amsmath,amssymb,floatfix]{revtex4}

\usepackage{graphicx}
\usepackage{bm}
\usepackage{pstricks}
\usepackage{xspace}
\usepackage{comment}
\usepackage[english]{babel}
\usepackage{amsfonts}%

\begin{document}

\title{Coarsening of ``clouds'' and dynamic scaling in a far-from-equilibrium model system}
\author{D.A. Adams, B. Schmittmann, and R.K.P. Zia}
\affiliation{Center for Stochastic Processes in Science and Engineering, \\
Department of Physics, Virginia Tech, Blacksburg, VA 24061-0435, USA}
\email{schmittm@vt.edu}
\date{\today}

\begin{abstract}
A two-dimensional lattice gas of two species, driven in opposite directions by an external
force, undergoes a jamming transition if the filling fraction is sufficiently high. 
Using Monte Carlo simulations, we investigate the growth of these jams (``clouds"),
as the system approaches a non-equilibrium steady state from a disordered initial state. 
We monitor the dynamic structure factor
$S(k_x,k_y;t)$ and find that the $k_x=0$ component exhibits dynamic scaling, of the form
$S(0,k_y;t)=t^\beta \tilde{S}(k_yt^\alpha)$. Over a significant range of times, we observe excellent
data collapse with $\alpha=1/2$ and $\beta=1$. The effects of varying filling fraction and
driving force are discussed. 
\end{abstract}

\maketitle

\section{Introduction}

\smallskip The study of phase separation and coarsening in systems
undergoing continuous or first-order phase transitions has a long history in
physics and materials science \cite{Langer,Gunton,Bray,Puri}. A model
system, like an Ising lattice gas, or a real alloy, like a mixture of tin
and lead, is prepared in a high-temperature state and then suddenly quenched
below its co-existence curve. As the system phase separates, its properties
are dominated by the morphology of growing single-phase domains. A
particularly interesting feature of many phase-ordering systems is dynamic
scaling: if space and time are appropriately rescaled, growing domains at
different times are found to be statistically self-similar, and the
characteristic domain size, $R(t)$, grows as a power of time, $t^{\alpha }$.
More detailed information is contained in the equal-time two-point
correlation function, or equivalently, the structure factor, which are
generalized homogeneous functions of space and time. For systems evolving
towards terminal \emph{equilibrium} states, the domain growth exponent $%
\alpha $ and the scaling behavior of correlation functions are fundamentally
well understood \cite{Bray}.

The situation is very different for many-body systems evolving towards 
\emph{non-equilibrium} steady states (NESS). Maintained far from equilibrium by
some external force, for example couplings to multiple energy or particle
reservoirs, these systems carry nonzero fluxes. As a result, their
stationary distributions lie outside the Boltzmann-Gibbs framework and are
known only for a few special cases. Yet, nonequilibrium systems occur
frequently in nature, particularly in many biological contexts. Not
surprisingly, they display much richer behaviors than systems in thermal
equilibrium \cite{SZ,Mukamel}, including a variety of pattern forming
instabilities and first-order phase transitions, controlled by the external
drive rather than a temperature variable. However, rather little is known
about coarsening phenomena in such systems. Given that the underlying
dynamics violates a very fundamental symmetry of equilibrium systems,
namely, detailed balance, it is not immediately obvious whether features
such as dynamic scaling or power law growth will persist when systems evolve
towards terminal states which fall into the NESS class.

As a first step towards a better understanding of coarsening in such
systems, it is instructive to investigate a few simple models, in the hope
that these will generate insights from which a more general theory can be
built. Looking for candidates which fall into the NESS class, which are well
characterized in other sectors of their phase diagram and exhibit
coarsening in some parameter regime, we are naturally led to driven
diffusive systems \cite{KLS,SZ}. These systems involve one, or several,
species of particles, diffusing on a lattice subject to a differential bias
and short-range interactions. Both the prototype, first introduced \cite{KLS}
as a deceptively trivial modification of the Ising lattice gas, and its
variants display many surprising and counterintuitive phenomena \cite{SZ}. A
particularly interesting modification involves models with two particle
species driven in opposite directions \cite{Hwang,Vilfan,Thies1} where
``jamming'' transitions emerge from biased diffusion alone.

Let us very briefly survey earlier studies of domain growth and dynamic
scaling in driven diffusive systems. The prototype model, an Ising-like
lattice gas in which the particles are ``charged'' and driven by an
external ``electric'' field $E$, sustains a nontrivial particle current on a
fully periodic lattice. Still, the order-disorder transition of the undriven
system survives, separating a disordered phase from a low-temperature phase
which phase-separates into high- and low-density strips, aligned with the
drive. If the system is quenched from a typical high-temperature state into
the phase-separated sector of the phase diagram, coarsening of single-phase
domains occurs \cite{ALLZ,LKM}. Some interesting morphological discrepancies
between simulation data and results from a continuum theory \cite{ALLZ} were
eventually resolved \cite{RY}. Turning to two-species models, the onset of
jamming separates a homogeneous, high-current phase from a spatially
inhomogeneous, low-current phase. As in the single species case, the jams
take the form of strips of high particle density, but these are now aligned 
\emph{transverse} to the field direction. At the late stages of the approach
to the steady state, the system typically exhibits several strips which
coarsen until only a single strip remains in the long-time limit.

Earlier work on dynamic properties has mostly focused on these late stages.
Since the strips are (on average) uniform in the transverse direction, they
are quite well described by a set of mean-field equations, in one space
dimension and time \cite{Vilfan,Thies2}. If the excluded volume constraint is
enforced rigorously, so that the particles are not allowed to swap places,
the strips coarsen logarithmically slowly \cite{KR,Thies2}. Another group of
studies investigates systems where the microscopic dynamics is already
restricted to one \cite{God,Gro}, or quasi-one \cite{Met,Geo}, dimension.
For interesting behavior to occur, particle-particle (``charge'') exchanges
must be permitted, albeit with a small rate, compared to particle-hole
exchanges. Provided the model parameters are chosen appropriately, compact
particle clusters form easily, and coarsen until one a single large cluster
remains. By virtue of the charge exchange process, power law growth
dominates here.

\begin{figure}[!t]
\includegraphics{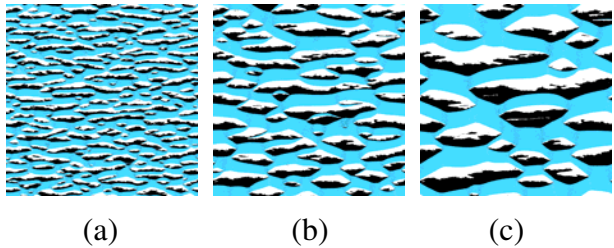}
\vspace{3.2cm}
\caption{Snapshots of an $800 \times 800$
system, at (a) $t=1024$, (b) $4096$, and (c) $16384$,
in units of MCS. $E=10$.
Positive (negative) particles are black (white); holes are blue.}
\label{three_configs}
\end{figure}

In this article, we present the first study of \emph{fully two-dimensional}
coarsening in a \emph{two-species} model with a strict excluded volume
constraint. Starting from an initially disordered configuration, the system
parameters (density, bias) are chosen so as to favor a jammed phase. Almost
immediately, small ``clouds'' (Fig.~\ref{three_configs}) of locally jammed particles form. The larger
clouds then grow, at the expense of the smaller ones, until a large cloud
percolates along the transverse direction, forming a strip. Eventually, several
strips emerge and compete with one another, on much slower time scales. 
We focus on the multi-cloud regime, long before the late-stage strip coarsening 
regime sets in. We monitor the equal-time
structure factor $S(\mathbf{k},t)$, as a function of wave vector $\mathbf{k}$
and time $t$, averaged over initial conditions and system histories. A range
of system sizes, densities, and $E$-values are studied. Since the field
selects a specific direction, the $y$-axis, it is not surprising that the
structure factors are anisotropic, in $k_{x}$ and $k_{y}$. More remarkably,
we find that the system exhibits good dynamic scaling in $k_{y}$ and $t$,
provided $k_{x}$ is fixed at $k_{x}=0$.
Assuming the scaling form $%
S(0,k_{y},t)=t^{\beta }\tilde{S}(k_{y}/t^{\alpha })$, the scaling exponents
are found to be $\alpha =1/2$ and $\beta =1$. For nonzero values of $k_{x}$,
or in the full $(\mathbf{k},t)$ domain, we have not been able to achieve
good data collapse.

This paper is organized as follows. We first present the model, a set of
diagnostic observables, and some technical details of the simulations. Next,
we discuss our simulation results and evidence for dynamic scaling. 
We conclude with some comments and open questions.\smallskip

\section{The model and its observables}

Our model is defined on a two-dimensional square lattice of size $%
L_{x}\times $ $L_{y}$ with fully periodic boundary conditions. Two species
of particles, referred to as ``positive'' and ``negative'', reside on the
sites of the lattice, subject to an excluded volume constraint. Hence, a
given configuration of the system can be labelled by a set of occupation
variables, $\sigma (\mathbf{r})$, taking the values $0$, $+1$, and $-1$ if the site
$\mathbf{r}=(x,y)$
is empty or occupied by a positive or negative particle, respectively. The
particles experience no interactions, apart from respecting an excluded
volume constraint. For simplicity, we restrict ourselves to systems which
are neutral:\ $\sum_{\mathbf{r}}\sigma (\mathbf{r})=0$.\ For later reference, we also
define the particle (as opposed to charge) occupation 
$n(\mathbf{r})$ via \smallskip 
\begin{equation}
n(\mathbf{r})=\left| \sigma (\mathbf{r})\right|  \label{dens}
\end{equation}%
so that the total particle density (``mass'') $m$ is given by $m =\left(
L_{x}L_{y}\right) ^{-1} \sum_{\mathbf{r}}n(\mathbf{r})$.

In the absence of the driving force, the particles perform simple diffusion,
i.e., jump with equal probability to a randomly selected nearest-neighbor
site, provided it is unoccupied. As a result, there is no net current (of
either mass or ``charge'')\ through the system, and the steady state is
spatially uniform. In contrast, an ``electric'' field, applied in the
positive $y$-direction, biases positive and negative particles in opposite
directions. In our simulations, a bond is selected at random and the
occupancies of the two associated sites are checked. If the bond carries a
particle-hole pair, an exchange will always be made if this results in a
positive (negative) particle moving in the transverse or positive
(negative)\ $y$-direction; otherwise, the exchange is attempted with rate $%
\exp (-E)$. Clearly, this dynamics is translation-invariant
and invariant under charge-parity transformation ($\sigma
\rightarrow -\sigma $, $y\rightarrow -y$). We use a random sequential
dynamics, with one Monte Carlo step (MCS) corresponding to $L_{x}\times
L_{y} $ update attempts. All runs start from a random initial condition.

The system sizes studied ranged from $100\times 100$ to $3200\times 3200$.
The density varied from $m =0.3$ to $m =0.7$. We also considered
different values for the probability for a particle to move backwards. Our
reference system, for which the largest data set was collected, is an $%
800\times 800$ lattice, with $m =0.5$ and $E=10$. The latter gives 
a probability of $4.5\times 10^{-5}$ for backward jumps which
is zero for all practical purposes. Runs lasted at least $8196=2^{13}$ MCS,
and data are typically averaged over $1,000$ runs, unless stated otherwise.
Time is measured in MCS.

The final stationary state of the system is well understood. For
sufficiently large particle density $m $ and field $E$, the system
displays a single strip of particles, transverse to the field direction 
\cite{Hwang}. 
In that fashion, translational symmetry is spontaneously broken.
The strip itself is charge-segregated, with
positive (negative) particles occupying sites with lower (higher) $y$%
-coordinates. The interior interface (separating positive from negative
particles) is glassy, due to the absence of any charge exchanges. In
contrast, the exterior interfaces (separating particles from holes) is quite
smooth, since its fluctuations are controlled by suppressed particle moves,
i.e., by the parameter $\exp (-E)$. This parameter also controls the density
of particles in the remainder of the system, reminiscent of a gas-liquid
interface under gravity. Due to the periodic boundary conditions, however, a
small current flows, even in the jammed phase, limited by $\exp (-EL_y)$. A
simple mean-field theory allows us to compute average density profiles,
currents, and the phase diagram, in good agreement with the simulations 
\cite{Hwang,Vilfan,Thies1}.

In the following, we will always choose system parameters such that the
system evolves towards an inhomogeneous, jammed final state. Starting from a
random initial disordered configuration, small jams of positive and negative
particles form very rapidly, due to local density fluctuations. Some of
these, typically the larger ones, will collect more particles and grow,
while others shrink and dissolve, as illustrated in 
Fig.~\ref{three_configs}. We refer to this stage as the coarsening of clouds, or
clusters. Eventually, first one and then several of the largest clouds will
span the lattice in the transverse direction, and the evolution is no longer
dominated by the coarsening of well-separated clouds. Now, multiple strips
compete for particles until only a single one remains, and the system has
reached its steady state.

In this study, the characteristic shapes and separations of the clusters are
of interest. Thanks to translational invariance, a suitable observable is
the equal-time structure factor, defined through the Fourier transform of
the local occupation,
\begin{equation}
\smallskip S(k_{x},k_{y};t)=\frac{1}{L_{x}L_{y}}\left\langle \left|
\sum_{x=0}^{L_{x}}\sum_{y=0}^{L_{y}}n(x,y;t)e^{i(k_{x}x+k_{y}y)}\right|
^{2}\right\rangle  \label{SF}
\end{equation}%
where $k_{x}=2\pi l/L_{x}$, $l=0,1,...,L_{x}-1$, and $k_{y}=2\pi j/L_{y}$, $%
j=0,1,...,L_{y}-1$. Here, $n(x,y;t)$ denotes the local occupation of site $%
(x,y)$ at Monte Carlo time $t$. The average $\left\langle ...
\right\rangle $ is taken over multiple runs, using configurations recorded
at the same Monte Carlo time. All initial conditions are random.

Let us first establish a few properties of this structure factor. The value
at the origin is easily found:
\begin{equation}
S(0,0)=\,\frac{1}{L_{x}L_{y}}<|\sum_{x,y}n(x,y)|^{2}>\,=%
m^{2} L_{x}L_{y}   \label{S(0,0)}
\end{equation}
Further, $S$ is related to the two-point correlation function, $G(x,y)\equiv
\left\langle n(x,y;t)n(0,0;t)\right\rangle $, via 
\begin{equation}
S(k_{x},k_{y};t)=\sum_{x,y}G(x,y)e^{i(k_{x}x+k_{y}y)}  \label{G}
\end{equation}
which also provides us with the sum rule 
\begin{equation}
\sum_{k_{x},k_{y}}S(k_{x},k_{y};t)=m L_{x}L_{y}  \label{sum}
\end{equation}
Finally, it is useful to evaluate $S$ for a few special cases, including the
initial and final configurations. Since the time argument is inessential
here, it will be suppressed for now. If the system is filled randomly with
particles, at density $m$, the structure factor is easily found to be 
\begin{eqnarray}
S(k_{x},k_{y}) &=&m (1-m )\left[ 1+O(1/(L_{x}L_{y}))\right] \nonumber \\
&+&m^{2}\left( L_{x}L_{y}\right) \delta _{k_{x},0}\delta _{k_{y},0}
\label{S_ini}
\end{eqnarray}
Clearly, $S(k_{x},k_{y})$ is uniform for all 
non-zero $\mathbf{k} \equiv (k_{x},k_{y})$. 
\smallskip

For comparison, we also evaluate the structure factor for a 
perfectly ordered single strip which reflects the stationary
state, modulo fluctuations:  
\begin{equation}
S(k_{x},k_{y})=\frac{L_{x}}{L_{y}}\delta _{k_{x},0}
\left[ 
\frac{ 1-\cos\left( m k_{y}L_{y}\right)}{1-\cos k_{y}}
\right] 
\label{S_fin}
\end{equation}%
These expressions provide a few benchmarks for the simulation data presented
below.

\section{Simulation results and tests for dynamic scaling}

\subsection{Unscaled structure factors.}

\begin{figure}[tbp]
\scalebox{0.6}{\includegraphics{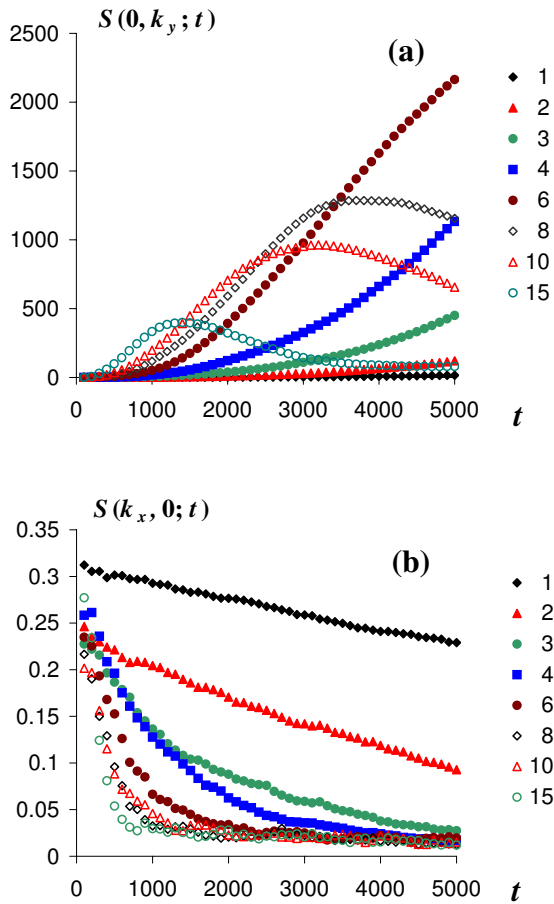}}
\caption{Unscaled structure factors $S(0,k_y;t)$ (a) and
$S(k_x,0;t)$ (b) for an $800 \times 800$
system, at $E=10$. The different curves correspond to different
values of $j$ (top) and $l$ (bottom), specified in the legend. 
The notation is that of Eq.~(\ref{SF}). Time is given in units of MCS.}
\label{unscaled-ver}
\end{figure}

In this section, we first present Monte Carlo data for raw (unscaled)
structure factors. We have collected data for a wide range of 
$\mathbf{k}$. Roughly speaking, the $\mathbf{k}$-value
of the peak \emph{position} reflects a characteristic spacing of the growing
clusters, while the peak \emph{width} carries information about
fluctuations. For illustration purposes, we show two projections here,
namely $S(0,k_{y};t)$ and $S(k_{x},0;t)$ (Fig.~\ref{unscaled-ver}). 
Plotted vs $k_{y}$, 
the data for $S(0,k_{y};t)$ show a distinct maximum which moves
towards smaller values of $k_{y}$ for later times. In contrast, $%
S(k_{x},0;t) $ is monotonically decreasing in $k_{x}$, for all $t$. A fuller
picture can be gleaned from contour plots of $S(k_{x},k_{y};t)$ which
indicate that, even for the earliest times considered ($t\leq 50$ MCS), the
maximum of $S$ is found on the $k_{x}=0$ axis. As time progresses, the peak
position shifts from larger values of $k_{y}$ to smaller ones, and the peak
height increases. These findings suggest that, as the first clouds emerge
from the fully disordered initial configurations, they quickly develop a
characteristic length scale in the field direction, but remain disordered in
the transverse direction.

\subsection{Scaled structure factors}

\begin{figure}[tbp]
\includegraphics{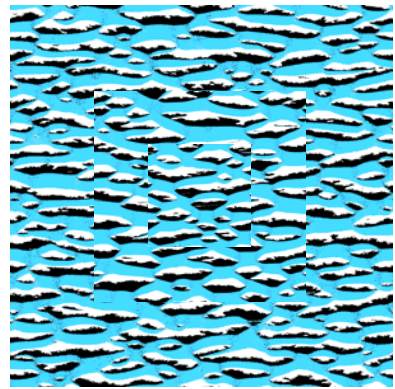}
\caption{Scaled configurations from Fig.~\ref{three_configs}. 
See text for details.}
\label{scaled_configs}
\end{figure}

\smallskip Snapshots of typical configurations at different times 
(Fig.~\ref{three_configs}) show clusters of particles (``clouds'') which grow in
both the parallel and the transverse directions. If a simple rescaling of
system size renders configurations, recorded at different times,
statistically similar, we can hope for dynamic scaling, as illustrated by
Fig.~\ref{scaled_configs}. After an appropriate rescaling of 
Figs.~\ref{three_configs}b and c, 
Fig.~\ref{three_configs}c is plotted inside Fig.~\ref{three_configs}b which is
plotted inside Fig.~\ref{three_configs}a. One has to take a very careful
look, if one wants to discern the internal boundaries (discontinuities)
between the three pictures. This illustrates - at a simple visual level -
how closely they resemble one another, after rescaling. However, our
visual ability to detect scaling is easily deceived and provides, at best, 
the \emph{motivation} for a more quantitative study. 

A \emph{quantitative} test of dynamic scaling requires a careful
analysis of the structure factors. 
Assuming that characteristic lengths in \emph{both} directions
increase as powers of time, but with possibly \emph{different} exponents due
to the anisotropy induced by the field, we first seek dynamic scaling in
the form 
\begin{equation}
S(k_{x},k_{y};t)\sim t^{\beta }f(k_{x}t^{\alpha _{1}},k_{y}t^{\alpha _{2}})
\label{dyn_scal}
\end{equation}%
where the $\sim $ indicates that we should expect this form to hold only for
certain ranges of time and wavevector. The sum rule, Eq. (\ref{sum}),
immediately leads to the exponent identity 
\begin{equation}
\alpha _{1}+\alpha _{2}=\beta  \label{exponents}
\end{equation}%
If dynamic scaling holds, one should be able to determine a set of scaling
exponents in such a way that structure factor data for different times and
wavevectors collapse onto a single curve if plotted according to Eq.~(\ref%
{dyn_scal}). However, we have not been able to achieve satisfactory data
collapse for this general form. Once again this suggests that there are no
characteristic \emph{transverse} length scales, associated with this growth
process. It also illustrates that merely visual tests of scaling, such as 
Fig.~\ref{scaled_configs}, must be treated with some caution.

\begin{figure}[tbp]
\scalebox{0.55}{\includegraphics{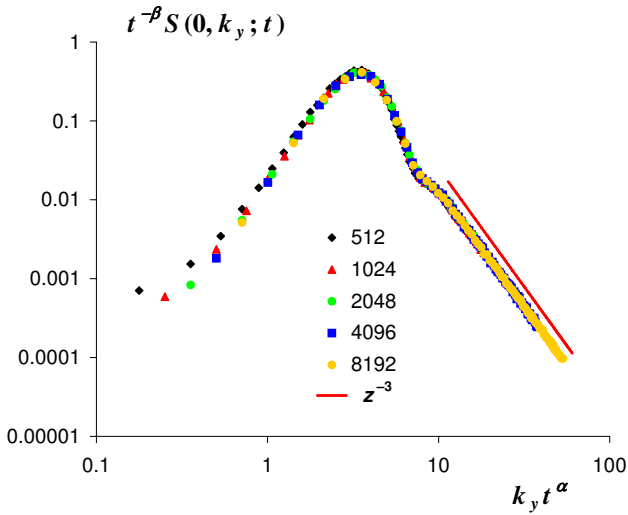}}
\caption{Scaling plot for $S(0,k_y;t)$ 
for an $800 \times 800$ system, at $m=0.5$ and $E=10$.
Five different times, ranging from $t=512$ 
to $t=8192$, in units of MCS, are shown. 
$\alpha=1/2$ and $\beta=1$. The solid line denotes a 
$z^{-3}$ power law. }
\label{scaled_0.5}
\end{figure}

Turning to the remnant structures in $S(0,k_{y};t)$, the data in 
Fig.~\ref{unscaled-ver} show a sequence of curves of
similar shapes, with the maximum shifting to smaller $k_y$ for later
times. Even if the general form, Eq. (\ref{dyn_scal}), is not obeyed, we can
explore the possibility of dynamic scaling in the reduced space $k_{x}=0$.
In the remainder of this article, we focus on tests of
\begin{equation}
S(0,k_{y};t)\sim t^{\beta }f(0,k_{y}t^{\alpha })  \label{zerokx}
\end{equation}%
Fig.~\ref{scaled_0.5} shows the scaling plot for a
half-filled system for times ranging from $t=2^{9}=512$ to $t=2^{13}=8192$.
We find excellent data collapse with the
scaling exponents $\alpha =0.50\pm 0.02$ and $\beta =1.00\pm 0.02$. Much
longer runs (with poorer statistics) show that the data continue to collapse
well, until at least $t \sim O(10^{6})$. Our value for $\alpha$,
the exponent controlling the characteristic spacing of domains, stands in
stark contrast to its counterpart for conserved coarsening in equilibrium
systems. There, it takes the value $1/3$, for a simple scalar density 
such as ours. 

The scaling function exhibits
Gaussian behavior near the maximum, and falls off as $z^{-3}$, where $%
z\equiv k_{y}t^{\alpha }$ is the scaling variable. This large 
$z$-behavior is highly reminiscent of the Porod tail \cite{Porod}, well 
known in the theory of domain growth in equilibrium systems. There,
it emerges from two essential features, namely, first, the presence of a 
single (isotropic) large length scale in the system, characterizing 
both the size and the separation of the coarsening domains, and second, 
the existence of 
microscopically sharp 
domain walls. Here, the situation is more
complex. While we do observe sharp domain walls between our clouds and
the surrounding (nearly) empty regions, our model is manifestly not isotropic.
What complicates the issue further is the absence of a characteristic
spacing in the direction transverse to the field. Clearly, 
a more careful study 
is required before the large $z$-behavior of our model can be traced
directly to a simple Porod law. 
As for the small $z$-behavior, we hesitate to offer any conclusions. Certainly, it does not appear to
follow the $k^4$ power law which would be expected for conserved coarsening
in equilibrium systems \cite{Yeung}.  

In the following, we probe the universality of the scaling exponents, as we change
system parameters such as the particle density or the driving force. First, we
consider the effect of system size. Since the scaling variable $z$
defines a characteristic length scale $\xi \propto t^{\alpha }$, it is
natural to expect a breakdown of scaling when $\xi $ becomes of the order of 
$L$, or when considering times $t\gtrsim L^{1/\alpha }$. Indeed, scaling
plots for a range of $L$, with $200\leq L\leq 566$ confirm this expectation 
very clearly. For example, in a $200\times 200$ system, the data for $%
t=10^{12}=4096$ already deviate noticeably from the scaling curve, whereas
for $L=566$, such deviations are not observed until $t=2^{15}=32768$.

\begin{figure*}[tbp]
\scalebox{0.7}{\includegraphics{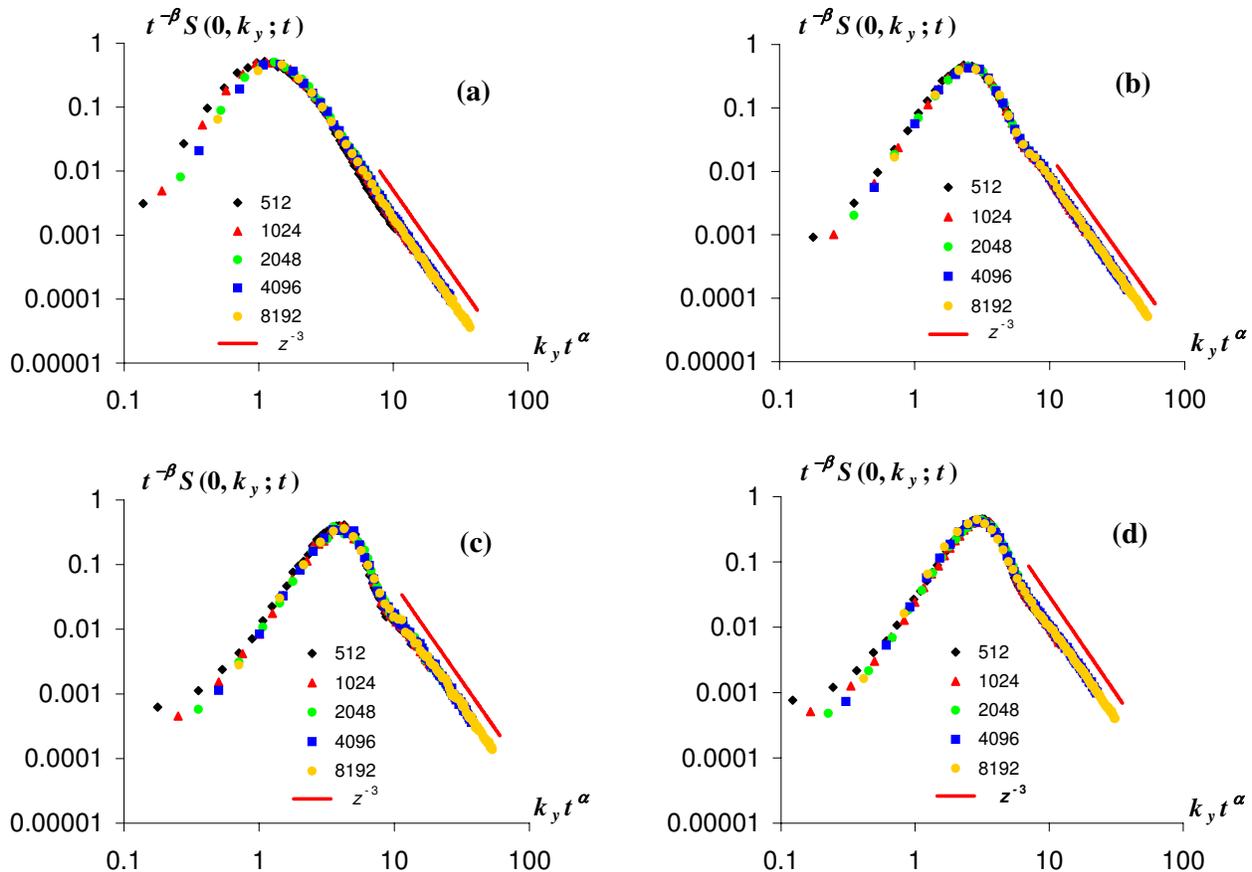}}
\vspace{0.7cm}
\caption{Scaling plots for $S(0,k_y;t)$ 
for an $800 \times 800$ system, at $E=10$.
Four different densities are shown, and the scaling exponents
are adjusted to give satisfactory data collapse: 
top left, $m=0.30$, $\alpha=0.46$, $\beta=0.98$; 
top right, $m=0.40$, $\alpha=0.50$, $\beta=1.00$; 
bottom left, $m=0.55$, $\alpha=0.50$, $\beta=1.00$; 
bottom right, $m=0.60$, $\alpha=0.44$, $\beta=0.95$. 
All errors are at most $5\%$. The solid lines denote 
$z^{-3}$ power laws. }
\label{scaled_other}
\end{figure*}

Next, we investigate the role of the overall particle density. For
coarsening in conserved equilibrium systems, it is well known that 
the scaling function depends on volume fraction of the minority phase; 
however, the scaling exponents describing the structure factor remain
unchanged \cite{Toral}. Here, the situation is much more dramatic. 
Using the scaling exponents $\alpha=1/2$ and $\beta=1$, the data
collapse for densities close to half-filling ($m=0.40$ and $0.55$) 
is still acceptable, but becomes progressively worse, for both 
larger ($m=0.70$) and smaller ($m=0.30$) densities. Better data
collapse can still be achieved, but at the price of modifying the scaling
exponents. Fig.~\ref{scaled_other} shows the scaled data, with 
appropriately adjusted values of $\alpha$ and $\beta$.
It is natural to assume that these values reflect \emph{effective}, 
rather than true asymptotic exponents. A better understanding of the 
scaling function would be necessary to disentangle its $m$-dependence 
from the overall scaling exponents.

We encounter a similar situation when considering the effect of the driving
force, $E$. We find good data collapse, with $\alpha =1/2$ and $\beta =1$, 
as long as the rate for a particle to move against its preferred direction, 
set by $\exp (-E)$, remains small. Once $\exp (-E)$ becomes comparable to 0.2, 
deviations from scaling become noticeable. More work will be required to 
shed light on these preliminary observations.

\section{\protect\smallskip Conclusions}

\smallskip To summarize, we have explored the possibility of dynamic scaling
in a two-dimensional driven lattice gas, involving two species of particles.
Positive and negative particles preferentially move in opposite directions
and form small jams, due to an excluded volume constraint. Above a certain
threshold density, these jams coarsen until a single strip of particles
spans the system in the transverse direction. For an extended period of
time, this coarsening process obeys dynamic scaling,  provided we focus on
characteristic length scales in the longitudinal direction. Monitoring a
structure factor, $S(0,k_{y};t)$, we find very good data collapse provided $%
t^{-\beta }S$ is plotted vs $k_{y}t^{a}$. At and near half-filling ($m=0.5)$
and for large driving force, we find $\alpha =0.50\pm 0.02$ and $\beta
=1.00\pm 0.02$. For smaller $E$ and densities further away from
half-filling, we believe that the scaling function acquires a dependence on $%
m$ and $E$. We note that we can still achieve reasonable data collapse with
the simple form given above, but only at the price of adjusting the
exponents $\alpha $ and $\beta $. We believe that these effective exponents
mask possibly significant modifications to the scaling function.

Naturally, a better analytic understanding of the exponents and of the 
scaling function would be desirable. It will be interesting to see 
what future studies in both simulations and analytics would reveal. The
observation that $\alpha $ is essentially $1/2$ points towards a diffusive
mechanism. Based on visual inspection alone, the clusters evolve by
exchanging particles with one another. If this process is truly random -
i.e., a cluster gains and loses particles with a fixed, constant rate, one
should indeed expect to find a diffusive growth of characteristic length
scales. Due to the drive, the particle exchange occurs predominantly between
clusters which are nearest neighbors in the transverse direction; hardly any
interactions occur between nearest neighbors in the transverse direction.
This may explain the absence of any apparent structures in $k_{x}$. Work is
in progress to analyze a well-established mean-field theory for this model,
in the hope of gaining a better understanding of exponents and scaling
function. If successful, it should also elucidate the deviations and 
similarities of our coarsening process with respect to those in 
equilibrium systems.

\emph{Acknowledgements}. We have benefitted from discussions with K.E. Bassler
and from the suggestions of a referee. This
work is supported in part by the NSF through DMR-0414122.

\end{document}